\documentstyle[osa]{revtex}
\input{psfig.sty}

\newcommand{\n}{\noindent}
\newcommand{\be}{\begin{equation}}
\newcommand{\ee}{\end{equation}}
\newcommand{\bea}{\begin{eqnarray}}
\newcommand{\eea}{\end{eqnarray}}
\newcommand{\mb}[1]{\mbox{\boldmath$#1$}}
\newcommand{\mr}[1]{\mathrm{#1}}            
\newcommand{\lb}{\left}                     
\newcommand{\rb}{\right}                    
                  
\newcommand{\gense}[1]{``#1''}        
\begin{document}

\begin{center}{\bf
Particle acceleration in three-dimensional tearing configurations
}
\end{center}

\begin{center}

{Christoph Nodes$^\dagger$, Guido T. Birk, Harald Lesch, and R\"udiger
Schopper\\
Institute for Astronomy and Astrophysics\\
University of Munich, Germany\\
also: Centre of Interdisciplinary Plasma Science, Garching, Germany\\
$^\dagger$ also: Max Planck Institute for Extraterrestrial Physics, Garching, 
Germany}
\end{center}

\begin{center}{\bf Abstract}\end{center}

\noindent

In three-dimensional electromagnetic configurations that result from unstable
resistive tearing modes particles can efficiently be accelerated to
relativistic energies. To prove this resistive magnetohydrodynamic simulations
are used as input configurations for successive test particle simulations.
The simulations show the capability of three-dimensional non-linearly evolved
tearing modes to accelerate particles perpendicular to the plane of the
reconnecting magnetic field components. The simulations differ considerably
from analytical approaches by involving a realistic three-dimensional electric
field with a non-homogenous component parallel to the current direction.  The
resulting particle spectra exhibit strong pitch-angle anisotropies.
Typically, about $5-8 \%$ of an initially Maxwellian distribution is
accelerated to the maximum energy levels given by the macroscopic generalized
electric potential structure. Results are shown for both, non-relativistic
particle acceleration that is of interest, e.g., in the context of auroral
arcs and solar flares, and relativistic particle energization that is
relevant, e.g., in the context of active galactic nuclei.
  

\newpage
\noindent
\baselineskip=5ex
\noindent {\bf I. INTRODUCTION}

\noindent
Magnetic reconnection is probably the most important mechanism of rapid
conversion of magnetic field energy to plasma heating and particle
acceleration$^{1-3}$ in fully and partially ionized plasmas. Cosmic plasma
phenomena as different as, e.g., flares in active galactic nuclei$^4$(AGN),
extragalactic jets$^{5, 6}$, solar flares$^7$ and discrete auroral arcs$^8$
probably show particle acceleration in reconnection regions.

The problem of particle acceleration is a long standing problem, since the
acceleration process depends critically on the local properties of the
reconnection region. The efficiency of the particle energization is a
function of the effective electric field force that acts on the particle.  In
two-dimensional reconnection configurations particle acceleration has been
extensively studied analytically [e.g., Ref. 9, 10] and numerically within the
framework of resistive [e.g., Ref. 11-13] and collisionless [e.g., Ref. 14-16]
reconnection.  Test particle simulations have proven the efficiency of
particle energization in turbulent reconnection [e.g., Ref. 17] These studies
have been performed for idealized electric and magnetic field configurations,
in particular homogeneous electric fields in the invariant direction, which
differ significantly from the three-dimensional configurations characterized
by localized field-aligned electric fields.

Test particle simulations of electron injection in high-beta plasmas have
shown that the investigation of more complex and therefore more realistic
three-dimensional field configurations is crucial for the understanding of
particle acceleration in reconnection regions$^{18}$.  Three-dimensional test
particle simulations of particle acceleration in reconnection regions that
result from shear flows have proven the short-comings of idealized analytical
two-dimensional treatments of the particle acceleration problem$^{19}$.

In the present contribution we consider the resistive tearing mode
instability, because it is considered as the generic spontaneous reconnection
process$^{20}$ [see also Ref. 2, 3 and references therein].  We present test
particle simulations that start from magnetohydrodynamic (MHD) configurations
resulting from the non-linear dynamics of tearing modes in three dimensions.

Our studies are relevant for particle acceleration in plasmas that can be
described as fluids. Processes that are of importance for particle
acceleration in collisionless plasmas as resonant nonlinear wave particle
interaction$^{21,22}$ are beyond the scope of our present approach.

The test particle simulations are carried out for different ratios of the
asymptotic equilibrium sheared magnetic field component to the field component
along the initial current direction.  For the initial particle distribution a
non-relativistic Maxwellian and a relativistic power law distribution are
chosen, respectively.  In the next section the initial non-linearly evolved
MHD configuration is introduced.  In Sec.~III the results of the test particle
simulations are presented for the non-relativistic and the relativistic cases.
Eventually, we discuss our findings in Sec.~IV.

\bigskip
\noindent{\bf II. THE MHD INITIAL CONFIGURATION}

\noindent
The MHD balance equations that govern the non-relativistic macroscopic
low-frequency dynamics read in a dimensionless
form
\be
{\partial  \rho\over \partial t} +\nabla\cdot(\rho{\bf v})=0
\label{1}
\ee
\be
{\partial  \rho {\bf v}\over \partial t}+\nabla\cdot(\rho{\bf v}{\bf v})
= -  \nabla p
+ \nabla \times {\bf B} \times {\bf B}
\label{2}\ee

\be
{{\partial p} \over {\partial t}} = - {\bf{v}} \cdot \nabla p -
\gamma p \nabla \cdot {\bf{v}} + (\gamma - 1) \eta 
(\nabla \times\bf{B})^2
\label{3}\ee

\be
{\partial {\bf B} \over \partial t} =
 \nabla \times ({\bf v}\times {\bf B}) -\nabla\times(\eta\nabla\times{\bf B})
\label{4}\ee
where $\rho$, ${\bf v}$, $p$, and ${\bf B}$ denote the mass density, bulk
velocity, thermal pressure, and the magnetic field.  By $\eta$ the normalized
diffusivity, i.e. the inverse magnetic Reynolds number, is denoted. All
quantities are made dimensionless by a typical mass density $\rho_0$, magnetic
field strength $B_0$, and length scale $L$, the Alfv\'en velocity
$v_A=B_0/\sqrt{4\pi\rho}$ and Alfv\'enic transit time $t_A=L/v_A$. Other
normalizing parameters follow in a generic way.

The balance equations are numerically integrated by means of a
three-dimensional finite differences code$^{23}$. The MHD simulations start
from an initial configuration given by
\be 
{\bf B}^{eq}= B_s{\rm tanh}(y){\bf e}_x +B_n {\bf e}_z\ \ \ ; 
\ \ \ \rho^{eq}=\rho_0 {\rm cosh}^{-2}(y)
\label{5}\ee
which is known as a Harris sheet equilibrium$^{24}$ with a leading magnetic
field component added along the current direction. It is assumed that the
ideal gas condition holds $p=nkT$ (where $n$, $T$, and $k$ are the particle
number density, temperature, and the Boltzmann constant).  The resistivity is
chosen to be localized around the midplane in the $z$-direction $\eta=\hat
\eta {\rm cosh}^{-2}((z-30)/3)$ with an amplitude of $\hat\eta=0.005$.  This
equilibrium, which can be regarded as an idealized standard current sheet
configuration, is perturbed by tearing eigenmodes
\bea
&&v_x = \hat v_x\rho_0 \ {\rm sin}\left(\frac{\pi x}{x_{\rm max}}\right)
\left(\left(1-\frac{\vert y\vert}{y_{\rm max}}\right)
\left(1-\frac{\vert y\vert}{\eta^{1/4}}\right)
     -\frac{\vert y\vert}{y_{\rm max}} \right)
e^{-\frac{\vert y\vert}{\eta^{1/4}}}\nonumber\\&&
v_y = -\hat v_y \rho_0 \ {\rm cos}\left(\frac{\pi x}{x_{\rm max}}\right)
\frac{\pi y}{x_{\rm max}}\left( 1-\frac{\vert y\vert}{y_{\rm max}}
\right) e^{-\frac{\vert y\vert}{\eta^{1/4}}}.
\label{6}\eea
The simulation runs were performed with the resolution of $77\times 77\times
63$ grid points in a numerical box given by $x\in[-25,25]$, $y\in[-10,10]$,
and $z\in[0,60]$ in normalized units. In the $x$- and $z$-direction an
equidistant discretization was chosen. The highest resolution of the
non-equidistant grid in the $y$-direction is 0.05. The non-linear dynamics of
the resistive tearing mode results in a magnetic field configuration with the
tearing-like $X$- and $O$-point structure in the $x$-$y$-plane.  The magnetic
field used as initial configuration for the first particle test simulations
(see Sec.~III) is shown in Fig.~1.
The respective three-dimensional electric field is shown in Fig.~2.
Obviously, a complex three-dimensional electromagnetic field configuration
appears which, differs significantly from analytical configurations used to
study particle acceleration. We emphasize that our simulations are
characterized by non-homogenous magnetic and electric field components
perpendicular to the reconnection flows, i.e. along the direction of the
initial current sheet. They make an important difference to any analytical
approach.

\bigskip
\noindent{\bf III. SIMULATIONS ON PARTICLE ACCELERATION}

\bigskip
\noindent
Since in many space and astrophysical circumstances unsaturated external shear
flows (stellar winds, stellar explosions, jets from quasars, etc...)  agitate
magnetic fields and drive the resistive tearing modes our MHD calculations are
used as the electromagnetic environment for the studies of electron
acceleration by means of test particle simulations. We simply want to know
how electric particles behave in the complex three-dimensional
electro-magnetic field configuration and neglect any back reaction of the
particles onto the field structure.

Thus, the electric field is derived only from the MHD quantities $\mb{B}$,
$\mb{v}$ and $\eta$ by means of Ohm's law
\be
  \mb{E} = - \frac{1}{c}\mb{v} \times \mb{B} + \eta \mb{\nabla} \times \mb{B}.
\ee
Since the data for the electric and magnetic field is only available on a
discrete three-dimensional grid the test particle code uses linear
interpolation to determine the field values for any location. We note that the
term \gense{test particle} simply means that the electromagnetic fields
produced by the particles are not changing the global fields, though the
effect of synchrotron radiation on the motion of the particles is taken into
account, i.e. the relativistic equations of motion have the form
\be
\frac{d\mb{p}}{dt} = q\left(\mb{E} 
+ \frac{1}{\gamma m c}\, \mb{p} \times \mb{B}\right) 
              + \mb{F}_{Rad}\;, \quad
\frac{d\mb{r}}{dt} = \frac{\mb{p}}{\gamma m},
\ee
where $m$ and $q$ denote the mass and charge of the particles, which are
electrons in our case and
\be
\gamma = \sqrt{1+\left(\frac{p}{mc}\right)^2}
\ee
is the Lorentz factor.

% Aufenthaltszeiten an der neutralen Linie?

The calculations are relativistic, including the energy losses via synchrotron
radiation and inverse Compton scattering by the term $\mb{F}_{Rad}$, which we
discuss in detail in Sect. IIIb. For the non-relativistic case of particle
acceleration, as considered in the following sub-section, the radiative losses
are negligible. The equations are numerically integrated by a Runge-Kutta
algorithm of fourth order with an adaptive stepsize control. As a result we
get the momentum and location at certain times of each particle in the given
ensemble.

\bigskip
\noindent{\bf IIIa. NON-RELATIVISTIC PARTICLE ACCELERATION}

\bigskip
\noindent
The choice of parameters for our studies of non-relativistic particle
acceleration is motivated by the phenomena of discrete auroral arcs 
for example. 
Reconnection may play an important role in the energization of auroral
particles in Birkeland current sheets$^{8, 25}$.
Our results are, however, to be understood as rather general ones for particle
acceleration in three-dimensional tearing configurations.  
Therefore, we present results from test particle simulation runs for three
different ratios of $B_n$ to $B_s$:
$B_n/B_s=1:1, 10:1,$ and $100:1$. The strength of the main magnetic field
component is chosen as $B_n^{1:1} \approx 10^{-3} \; \mr{Gauss}$. The
electric field components are $E_x \approx 10^{-5} B_n$ and $E_y
\approx E_z \approx 10^{-4} B_n$, respectively. 
It is a characteristic feature of 3D reconnection that the magnitude of the
electric field component directed along the initial current sheet is of the
order of the dominant convective electric field component.
The spatial parameters for the first three runs were adjusted to those of
auroral phenomena, i.e. the size of the box is $2000 \times 800 \times
2400\;\mr{km}$. The initial position of the particles is uniformally
distributed in the (x,y,0)-plane at the lower $z$ end of the box.
For the particle distribution we choose a Maxwellian distribution
with a corresponding temperature of $k_B T\approx 1\;\mr{eV}$ and
a shift of the momentum in the $z$-direction of a few percent of the
thermal energy, thus the distribution is in fact slightly anisotropic.
We consider the state when all electrons have left the computational box as
the final state.

The trajectories of all electrons for each run are shown in Fig.~3-5 as a
projection on the $y$-$z$-plane. The gyration motions of the electrons occur
on a much smaller scale so that this motion cannot be identified, but it is
fully resolved in the simulation.
The reconnection zone is placed between $z\approx 1000$ and $z \approx 1500$
where the $\mb{E}\times\mb{B}$-drift forces the electrons to move towards the
center of the region. Those electrons which start at coordinates with a
$y$-value close to zero are most efficiently accelerated. 
They leave the box at high $z$-values while the others leave the box at the
$x$ and $y$ boundaries following the magnetic field lines. The acceleration
takes place on very short timescales of a few $10^{-1}$ seconds.

Fig.~6-8 present the initial and final energy spectra of the
electrons. Obviously, most of the electrons keep their initial Maxwellian
distribution. The injected test-particles are not noticeably heated. Some
fraction of the electrons are accelerated and show a power law with an average
index of $1.1$ to $1.5$. The main recognizable result is the pile-up of
electrons at the high energy tail of the distribution, which shows up in all
simulations. The heights of the peaks, i.e. the number of accelerated
electrons, increase with decreasing $B_n/B_s$ (Fig.~6 to Fig.~8), while the
maximum energization is approximately constant.

In Fig.~9-12 the relative momentum distribution of the test particles of the
first run is displayed. The $p_x-p_z$ cut of the phase space shows that the
accelerated particles have comparable momentum gains in $x$- and $z$-direction
(Fig.~9-10), whereas there is almost no momentum gain in the $y$-direction
(Fig.~11-12). For the cases with smaller ratios of $B_n/B_s$ the acceleration
is almost exclusively along the current sheet.

The pitch angle distribution (Fig.~13) shows a clear anisotropy in the final
state, i.e. there is a strong peak of the distribution function at
$\cos(\theta)=1$ along with a depression in between $0<\cos(\theta)<0.9$. This
might be explained by the fact that the electrons get accelerated by the
component of the electric field parallel to the magnetic field,
$E_{\|}$. Thus, they only gain momentum parallel to the magnetic field while
the perpendicular component $p_{\perp}$ remains constant (just like in the
homogeneous case). Consequently, the momentum vector is predominantly directed
parallel to the magnetic field.
We would like to emphasize that such a pitch angle anisotropy is
characteristic for the so-called inverted V events detected by satellites
below the main auroral acceleration regions$^{26}$. Thus, our results
corroborate the current striation model for auroral acceleration in which the
tearing instability plays a central role$^8$.

Fig.~14 shows the initial and final pitch angle distribution for the most
energetic particles of the first run with $B_n/B_s = 1:1$. This plot clearly
indicates that the pitch angles of the most energized particles depend on
their initial pitch angle as it is expected for a non stochastic acceleration
process.

We emphasize that different from two-dimensional studies$^{27}$ a significant
fraction of the particles is not trapped in the O-point regions and thus
efficiently accelerated.

\bigskip
\noindent{\bf IIIb. RELATIVISTIC PARTICLE ACCELERATION}

\bigskip
\noindent
In this section we discuss simulations in a parameter range which is typical
for the central regions of active galactic nuclei. Such objects like quasars
are accreting massive black holes surrounded by magnetized accretion disks
from which relativistic plasma jets emanate. They exhibit very rapid flares in
almost all bands of the electromagnetic spectrum, especially in the
TeV-range$^{28}$. The spectra are power laws, i.e. the radiation is of
nonthermal origin, for example synchrotron radiation and/or inverse Compton
scattering. The time scales for some these flares are down to 30 min. In other
words, within that short time scale a significant number of particles has to
be accelerated very efficiently to provide the observed radiation
fluxes. Since we know that these objects contain differentially rotating
magnetized gas disks surrounded by magnetized coronae we think that magnetic
reconnection may provide the required fast particle acceleration$^{29}$. As
mentioned above the tearing mode is a generic configuration for reconnection,
thus we consider acceleration under circumstances typical for the central
light year of an active galactic nucleus. We note that our simulations do not
consider an individual object or flare event but concentrate on the general
mechanism for energetization of relativistic particles in magnetized
plasmas. Especially we do not consider the plasma at some specific distance
from the central black hole, rather we simulate the part of a plasma volume
with a magnetic field typical for the central $10^{18}\, cm$ above the
magnetized accretion disk and close to the magnetized emanating jet.
Accordingly, the size of the computational box was chosen as $(100 \times 40
\times 120)\times 10^{13}\;\mr{cm}$. The main component of the magnetic field
as well as the asymptotic shear component was scaled to $B_n \approx
0.1\;\mr{Gauss}$, i.e. the ratio for all relativistic runs was $B_n/B_s =
1:1$. The associated electric field reads $E_x \approx 10^{-3}, E_y \approx
10^{-2}, $ and $E_z \approx 10^{-3}$ in units of $B_n$.

We show results of three different runs. One run starts with a relativistic
Maxwellian distribution (with $k T/mc^2 =50$) and in the other runs we use
initial power law energy distributions for the injected electrons with Lorentz
factors ranging from $10^4$ up to $2\cdot 10^6$ and $10^2$ up to $2\cdot
10^4$, respectively. A power law distribution mimics a pre-accelerated
electron population in an active galactic nucleus.

In the considered high energy ranges (GeV to TeV) the losses due to
synchrotron radiation and inverse Compton scattering certainly play a role,
thus the term
\begin{eqnarray}
  \label{frad}
  \mb{F}_{Rad} 
  & \approx & \frac{2}{3} \frac{q^4}{m^2 c^4} 
  \lb\{ \mb{E} \times \mb{B} + 
  \frac{1}{c} \mb{B} \times \lb(\mb{B} \times \mb{v}\rb) + 
  \frac{1}{c} \mb{E} \lb(\mb{E\cdot v}\rb) \rb\}   \nonumber \\
  & - & \frac{2}{3} \frac{q^4}{m^2 c^5} \gamma^2 \mb{v} 
  \lb\{ \frac{16\pi}{3} U_{Ph} + \lb(\mb{E} + 
  \frac{1}{c} \mb{v} \times \mb{B}\rb)^2 - 
  \frac{1}{c^2} \lb(\mb{E\cdot v}\rb)^2 \rb\} 
\end{eqnarray}
now becomes important for the motion of the electrons (see Eq.(8)). Expression
(\ref{frad}) is an approximated form of the damping force due to radiation
given by Landau and Lifshitz$^{30}$ which we used in our code. $U_{Ph}$
denotes the photon energy density, which has to be considered for inverse
Compton scattering. For AGN coronae a typical value is $U_{Ph}\approx
10^{-4}\;\mr{erg/cm^3}$.

The resulting synchrotron spectra can be calculated as the sum of total power
emitted by each individual electron. The latter is given by$^{31}$
\begin{equation}
P\left(\frac{\omega}{\omega_c}\right) = \frac{\sqrt{3}}{2 \pi} \frac{e^3 B}{mc^2}
  \sin \theta \frac{\omega}{\omega_c} \int_{\frac{\omega}{\omega_c}}^\infty 
  K_\frac{5}{3}(\xi) d\xi,
\end{equation}
where $\omega_c$ is the critical frequency, $\theta$ is the pitch angle and
$K_\frac{5}{3}$ is the modified Bessel function.

Let us first discuss the case with an initial power law with an energy range
from $\gamma = 10^4-2\cdot10^6$. The initial and final energy spectra are
presented in Fig.~15. Quite a large fraction of the electrons get decelerated
below $\gamma\sim 10^4$ whereas some reach energies up to $\gamma\sim 10^7$.
The deceleration of particles is caused by the electromagnetic forces and
associated radiation losses they experience along their trajectories.
Different from the accelerated particles they do not enter the inner
reconnection region where the dominant electric field component is directed
parallel to their guiding center motion. Their enhanced perpendicular momentum
gives rise to efficient radiative losses.

The distribution also shows this beam-like built-up of electrons at higher
energies which was observed in the low energy simulations. Unlike the
non-relativistic simulations, here the electrons get somewhat decelerated
after leaving the reconnection region due to the radiation losses, but still
have much higher energies in the end than before.

The deceleration of electrons is also visible in the radiation spectrum of the
synchrotron emission since the initial spectrum, or radiation power
distribution, reaches up to higher frequencies than the final spectrum
(Fig.~16). On the other hand, the accelerated particles ($\gamma >10^6$) have
a very small pitch angle and therefore do not significantly contribute to the
synchrotron emission within our simulation volume. Interesting enough such an
anisotropic particle distribution is able to transport energy to regions far
away from the acceleration site, since the particles do not lose energy by
synchrotron radiation. Such a behavior has been demanded by Ghisellini$^{32}$
for the X-ray and gamma-ray bursting blazars. There must be some energy
transport mechanism which is dissipationless up to large distances from the
acceleration site, and the appearance of an anisotropic ultrahigh energy
component would present such a mechanism $^{33}$.

The selection which electrons are to be accelerated or decelerated is
certainly due to a geometric effect, i.e. only those electrons which have an
appropriate initial position experience a strong acceleration. This is visible
in Fig.~17-21 in which the trajectories of these two classes of particles are
presented. Obviously, only those electrons are most efficiently accelerated
which are injected very close to the quasi-separatrix layers. The
separatrix-like projection onto the $x$-$y$-plane does not necessarily
indicate regions of different magnetic topology in the considered
three-dimensional configuration. The trajectories of accelerated and
decelerated particles are somehow complementary to each other. An interesting
feature is the sickle-shaped form of the trajectories in the
z-y-projection. It seems as if the reconnection zone acts like a defocusing
lens to the accelerated electrons. The decelerated electrons do not move along
reconnecting magnetic field lines that cross the initial neutral layer.

In the second run, initialized with an energy range from $\gamma = 10^2-2\cdot
10^4$, the energy and synchrotron spectra are shown in Fig.~22 and 23. The
pile-up at $\gamma \sim 8\cdot 10^6$ is less pronounced, but still
present. The radiation spectrum clearly exhibits a broken power law. The low
energy part of the power law is mainly due to the initially injected
particles, whereas the high energy part is caused by the accelerated particle
population.  The particle trajectories are qualitatively similar to the
previously discussed run. The same statement holds for the case of the initial
relativistic Maxwellian particle distribution. For this case also efficient
acceleration occurs (Fig.~24, 25) but there is no pile-up at high
energies. The radiation exhibits a rather flat spectrum at higher
energies. This is of particular interest in the context of observations
indicating flat nonthermal spectra of blazars$^{34,35}$ like Mrk 421 and Mrk
501.

\bigskip\noindent{\bf IV. DISCUSSION}

\bigskip\noindent 
Particle acceleration in magnetized collisionless plasmas is a central issue
of high energy astrophysics and space physics. A large number of observation
in all bands of the electromagnetic spectrum demand for a fast and efficient
production of accelerated particles within magnetized plasmas which are
agitated by unsaturated external forces like differential rotation, jets,
stellar winds and explosive events.  Such forces twist, shear and stretch
field lines into configurations in which magnetic reconnection plays a
dominant role either as relaxation mechanisms and/or as acceleration
process. The tearing instability is known to be a generic configuration in
such reconnection sites.

Thus, we studied the acceleration of non-relativistic and relativistic
electron populations in three-dimensional tearing configurations. By means of
test particle simulations performed within non-linearly evolved
electromagnetic fields modeled by MHD simulations we could show that particles
are effectively accelerated. Our simulations include nonthermal radiative
losses like synchrotron radiation. Starting from a shifted Maxwellian
distribution we carried out test particle simulations for three different
ratios of the magnitudes of the initial magnetic field components parallel and
perpendicular to the current sheet. For example, we have chosen physical
parameters that are relevant for auroral acceleration zones. Our findings are,
however, of general interest for particle acceleration in reconnecting current
sheets.  The degree of the pitch angle anisotropy of the accelerated particles
depends on the strength of the guiding magnetic field component along the
current sheet. A significant fraction of the initially injected electrons
gains momentum. The level of energization is limited by the overall
generalized electric potential $U=\int ds E_{\vert\vert}$.

The relativistic studies dealt with the fate of electrons that enter the
reconnection region with a power law distribution. Such a situation is
characteristic for flares in active galactic nuclei. Some fraction of the
pre-accelerated particles is fast and efficiently energized in the
three-dimensional tearing configuration. As in the runs that start from a
Maxwellian distribution a high-energy bump forms in the final distribution.

Our simulations prove the capability of magnetic reconnection to energize an
electron population of a magnetized plasma under the influence of external
electromagnetic fields onto which the accelerated leptons have no back
reaction. For example, close to a black hole the differential rotation of the
surrounding accretion disk which shears magnetic field lines is only
determined by the mass of a central object but not influenced by the
dissipation of magnetic energy in small current sheets.  In other words, the
external forces which stretch, fold and twist the magnetic field lines act on
a much larger spatial scale than the size of the forming current sheets. This
behavior is similar to what is known from turbulent plasmas, in which the
energy input on large spatial scales is finally dissipated on considerably
smaller length scales. Thus, the energy sources and energy sinks are
completely disconnected from each other. Furthermore, the enormous strength of
such global forces like differential rotation substantiates our test particle
approach, which implicitly assumes no back reaction of the particles onto the
electrodynamic structure of the plasma.

Our present studies do not include turbulent electromagnetic fields. Numerical
studies of particle acceleration in MHD turbulent reconnection regions seem to
be a promising task for the future.
\newpage
\begin{center} {\bf REFERENCES}
\end{center}
\newcounter{no}
\begin{list}
{[\arabic{no}]}{\usecounter{no}}

\item
R.D. Bentley and J.T. Mariska (eds.), 
{\it Magnetic Reconnection in the Solar Atmosphere}, ASP Conf. Ser. 
{\bf 111} (1996).

\item
E.R. Priest and T. Forbes,
{\it Magnetic reconnection: MHD theory and applications} 
(Cambridge University Press, New York, 2000).

\item
D. Biskamp, {\it Magnetic Reconnection in Plasmas} (Cambridge 
University Press, Cambridge, 2000).

\item
H. Lesch and G.T. Birk, Astron. Astrophys. {\bf 324}, 461 (1997).

%5
\item
 E.G. Blackman, Astrophys. J. {\bf 456}, L87 (1996).

\item
H. Lesch and G.T. Birk, Astrophys. J. {\bf 499} 167 (1998).

\item
E.R. Priest and T. Forbes, Astron. Astrophys. Rev. {\bf 10}, 313 (2000).

\item
A. Otto and G.T. Birk, Geophys. Res. Lett. {\bf 20}, 2833 (1993).

\item
J.S. Wagner, P.C. Gray, J.R. Kan, T. Tajima, and S.-I. Akasofu,
Planet. Space Sci. {\bf 4}, 391 (1981).

%10
\item
Y.~Litvinenko, Astrophys.~J. {\bf 462}, 997 (1996).

\item
D.L. Bruhwiler and E.G. Zweibel, J. Geophys. Res. {\bf 97}, 10825 (1992).

\item
W.~Moses, J.M.~Finn, and K.M.~Ling,
J.~Geophys.~Res. {\bf 98}, 4013 (1993)

\item
B. Kliem, Astrophys. J. Supp. {\bf 90}, 719 (1994).

\item
R. Horiuchi and T. Sato, Phys. Plasma {\bf 4}, 277 (1997).

%15
\item
G.E. Vekstein and P.K. Browning, Phys. Plasma {\bf 4}, 2261 (1997).

\item
Y.E. Litvinenko, Phys. Plasma {\bf 4}, 3439 (1997).

\item
J. Ambrosiano, W.H. Mattheus, M.L. Goldstein and
D. Plante, J. Geophys. Res. {\bf 93}, 14383 (1988).

\item
J. Birn, M.F. Thomsen, J.E. Borovsky, G.D. Reeves,
D.J. McComas and R.D. Belian, J. Geophys. Res. {\bf 103}, 9235 (1998).

\item
R. Schopper, G.T. Birk, and H. Lesch, Phys. Plasmas {\bf 6}, 4318 (1999).

%20
\item
H.P. Furth, J. Killeen, and M.N. Rosenbluth, Phys. Fluids  {\bf 6}, 459 (1963).

\item
M. Hoshino, J. Arons, Y. A. Gallant, A. B. Langdon, Astrophys. J. {\bf 390},
454 (1992).

\item
D. B. Melrose, in {\it Plasma Astrophysics}, eds. J. G. Kirk, D. B. Melrose, and
E. R. Priest (Springer, Berlin, 1994) p. 113.

\item
A. Otto, Comp. Phys. Com. {\bf 59}, 185, (1990).

\item
E.G. Harris, Nuovo Cimento {\bf 23}, 115 (1962).

%25
\item
G.T. Birk and A. Otto, J. Atm. Sol.-Terr. Phys., {\bf 59}, 835, (1997).

\item
C.S. Lin and R.A. Hoffmann, J. Geophys. Res., {\bf 84}, 1514 (1979).

\item
J.S. Wagner, J.R. Kan and S.-I. Akasofu, J. Geophys. Res., {\bf 84},
891, (1979).

\item
J.A. Gaidos, C.W. Akerlof, S.D. Biller {\it et al.}, Nature, {\bf 383}, 319 (1996).

%\item S. Wagner and A. Witzel, Ann. Rev. Astron. Astrophys., {\bf 33},
%163 (1995).

\item
H. Lesch, {\it Plasmas in the Universe}, eds. B. Coppi, A. Ferrari, E. Sinodi 
(IOS Press, Amsterdam, 2000) p. 395.

\item
L. D. Landau and E. M. Lifshitz, 
{\it The Classical Theory of Fields} (Addison-Wesley, Reading MA., 1951).

\item
G. B. Rybicki and A. P. Lightman, 
{\it Radiative Processes in Astrophysics} (Wiley, New York, 1979) ch. 6.2.

\item
G. Ghisellini,
{\it BL Lac Phenomena}, Astronomical Society of the Pacific, 159, 311 (1999).

\item
A. Crusius-W\"atzel and H. Lesch, Astron. Astrophys., 299, 404 (1998).

\item
G. Tosti, {\it et al.}, Astron. Astrophys., 339, 41 (1998).

\item
P. G. Edwards, G. Giovannini, W. D. Cotton, L. Feretti, K. Fujisawa,
H. Hirabayashi, L. Lara, and T. Venturi, Publ. Astron. Soc. Japan, 52, 1015
(2000).

\end{list}

\newpage\n
{\bf FIGURE CAPTIONS}

\bigskip\n
Figure~1: The magnetic field configuration that results from the 
MHD simulation after 180 Alfv\'en times.
The upper plot shows magnetic field lines, the lower one shows an arrow
plot of the $x$- and $y$-components of the magnetic field at $z=30$.

\bigskip\n
Figure~2: The electric field configuration after $t=180\tau_A$. 
The upper plot shows electric field lines, the lower one shows an arrow
plot of the $y$- and $z$-components of the electric field at $x=3$.

\bigskip\n
Figure~3: The trajectories of all electrons of the first run ($B_n/B_s=1:1$)
as a projection on the $y-z$ plane.

\bigskip\n
Figure~4: The trajectories of all electrons of the second run ($B_n/B_s=10:1$)
as a projection on the $y-z$ plane.

\bigskip\n
Figure~5: The trajectories of all electrons of the third run ($B_n/B_s=100:1$)
as a projection on the $y-z$ plane.

\bigskip\n
Figure~6: The spectrum of kinetic energy in the first run. The relative number
frequency is plotted against the kinetic energy in units of $mc^2$. The dotted
line represents the initial distribution while the normal line is the final
distribution.

\bigskip\n
Figure~7: The spectrum of kinetic energy in the second run.

\bigskip\n
Figure~8: The spectrum of kinetic energy in the third run.

\bigskip\n
Figure~9: The initial momentum plotted in the $p_x-p_z$ plane of phase
space of the first run ($B_n/B_s=1:1$).

\bigskip\n
Figure~10: The gain of momentum in $z$ direction compared to that in $x$
direction. For most of the accelerated particles $p_z$ is only slightly higher
than $p_x$.

\bigskip\n
Figure~11: The initial momentum plotted in the $p_y-p_z$ plane of phase
space of the first run ($B_n/B_s=1:1$).

\bigskip\n
Figure~12: The gain of momentum in $z$ direction compared to that in $y$
direction. There is no noticeable gain in $y$ direction.

\bigskip\n
Figure~13: The pitch angle distribution at the beginning (dotted line) and at
the end (normal line) of the first simulation ($B_n/B_s=1:1$). The relative
number frequency is plotted against $\cos(\theta)$.

\bigskip\n
Figure~14: The pitch angle distribution of the electrons with the highest
energy gain. These electrons already start with very small pitch angles.

\bigskip\n
Figure~15: The energy spectrum of the first relativistic run. This simulation
started with a power law (dotted line) of 
$N(\gamma-1)d\gamma=(\gamma-1)^{-2}d\gamma$. The final spectrum (normal line)
shows a local maximum at $(\gamma-1 \sim 8\times10^6)$.

\bigskip\n
Figure~16: The synchrotron spectrum of the first relativistic run.

\bigskip\n
Figure~17: The trajectories of the relativistic run as a projection on the
$y-z$ plane.

\bigskip\n
Figure~18: The trajectories of the accelerated electrons as projection on the
$y-z$ plane.

\bigskip\n
Figure~19: The trajectories of the decelerated electrons as projection on the
$y-z$ plane.

\bigskip\n
Figure~20: The trajectories of the accelerated electrons as projection on the
$x-y$ plane.

\bigskip\n
Figure~21: The trajectories of the decelerated electrons as projection on the
$x-y$ plane.

\bigskip\n
Figure~22: The energy spectrum of the second relativistic run. This simulation
started with a power law (dotted line) of
$N(\gamma-1)d\gamma=(\gamma-1)^{-2}d\gamma$. The final spectrum (normal line)
shows a pile-up at $(\gamma-1 \sim 8\times10^6)$.

\bigskip\n
Figure~23: The synchrotron spectrum of the second relativistic run.

\bigskip\n 
Figure~24: The energy spectrum of the third relativistic run that started with
a power law (dotted line) of $N(\gamma-1)d\gamma=(\gamma-1)^{-2}d\gamma$.

\bigskip\n
Figure~25: The synchrotron spectrum of the third relativistic run.

\newpage

% MHD

\begin{figure}
  \psfig{file=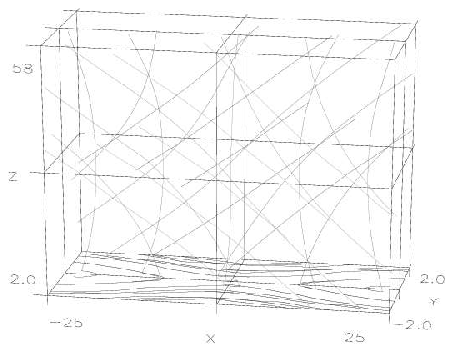,width=0.8\textwidth,angle=90}

  \psfig{file=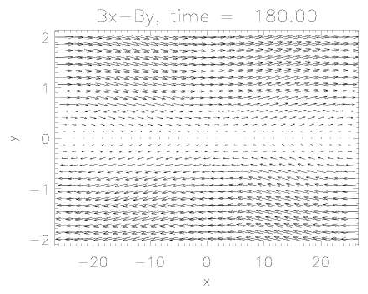,width=0.4\textwidth}
\caption[]{}
\end{figure}

\begin{figure}
  \psfig{file=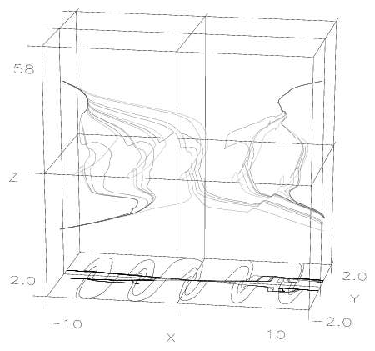,width=0.8\textwidth,angle=90}

  \psfig{file=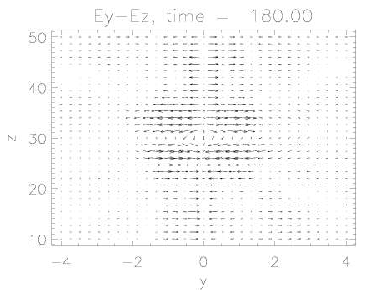,width=0.4\textwidth}
\caption[]{}
\end{figure}

% trajectories

\begin{figure}
    \psfig{file=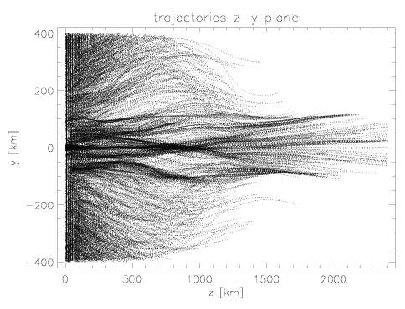,width=0.8\textwidth}
\caption[]{}
\end{figure}

\begin{figure}
    \psfig{file=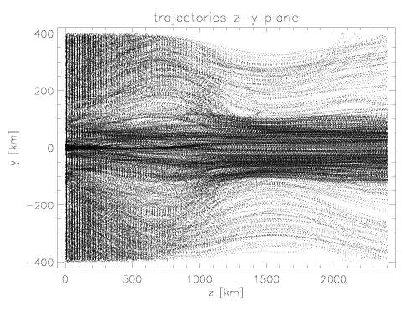,width=0.8\textwidth}
\caption[]{}
\end{figure}

\begin{figure}
    \psfig{file=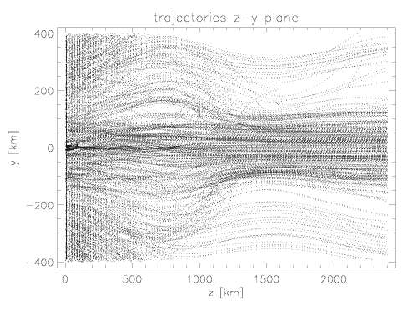,width=0.8\textwidth}
\caption[]{}
\end{figure}
\clearpage

% energy spectra

\begin{figure}
    \psfig{file=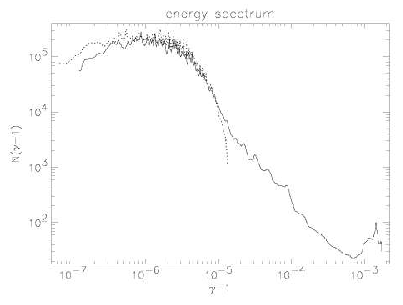,width=0.8\textwidth}
\caption[]{}
\end{figure}

\begin{figure}
    \psfig{file=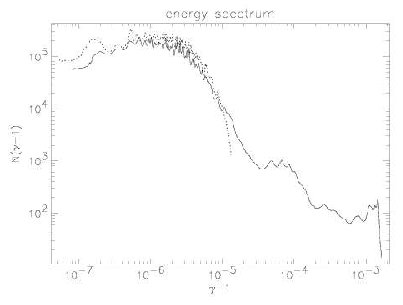,width=0.8\textwidth}
\caption[]{}
\end{figure}

\begin{figure}
    \psfig{file=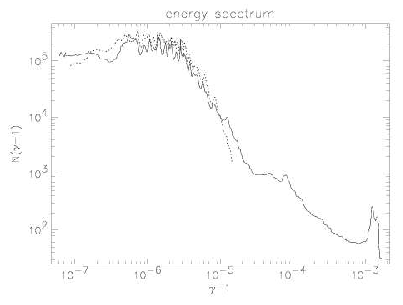,width=0.8\textwidth}
\caption[]{}
\end{figure}
\clearpage

% misc

\begin{figure}
    \psfig{file=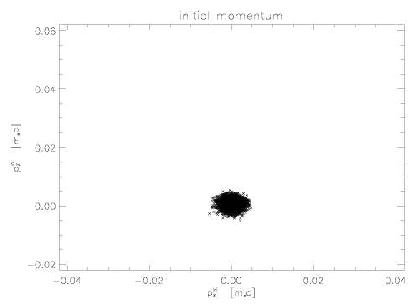,width=0.8\textwidth}
\caption[]{}
\end{figure}
\begin{figure}
    \psfig{file=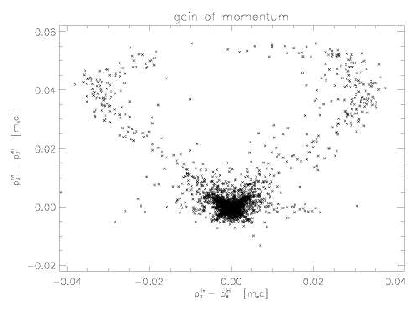,width=0.8\textwidth}
\caption[]{}
\end{figure}

\begin{figure}
    \psfig{file=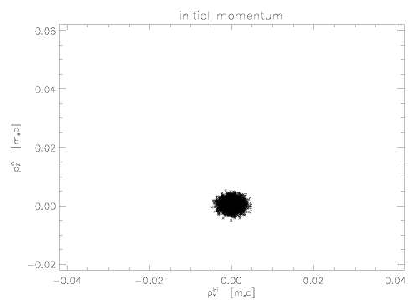,width=0.8\textwidth}
\caption[]{}
\end{figure}
\begin{figure}
    \psfig{file=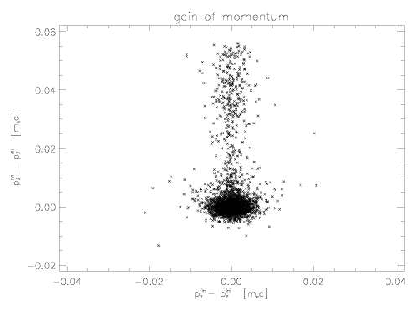,width=0.8\textwidth}
\caption[]{}
\end{figure}

\begin{figure}
    \psfig{file=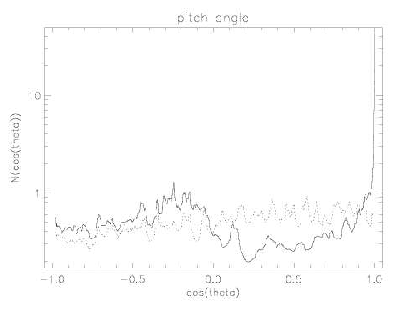,width=0.8\textwidth}
\caption[]{}
\end{figure}
\begin{figure}
    \psfig{file=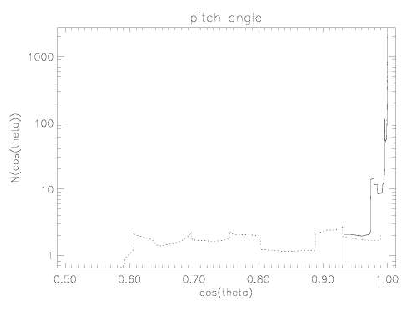,width=0.8\textwidth}
\caption[]{}
\end{figure}
\clearpage

% tearing20

% energetics

\begin{figure}
    \psfig{file=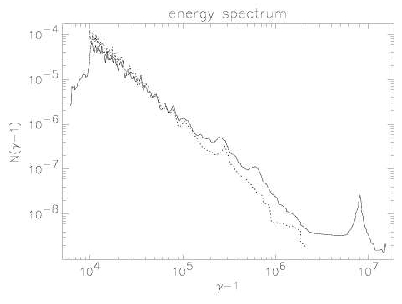,width=0.8\textwidth}
\caption[]{}
\end{figure}

\begin{figure}
    \psfig{file=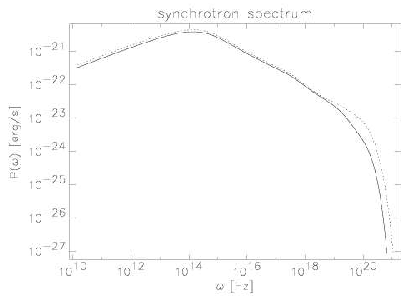,width=0.8\textwidth}
\caption[]{}
\end{figure}
\clearpage

\begin{figure}
    \psfig{file=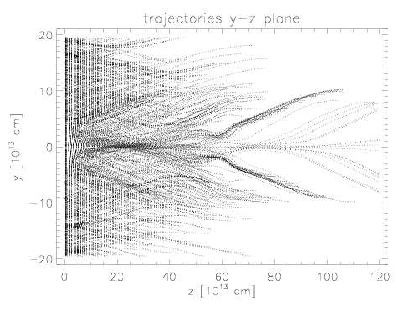,width=0.8\textwidth}
\caption[]{}
\end{figure}
\clearpage

\begin{figure}
    \psfig{file=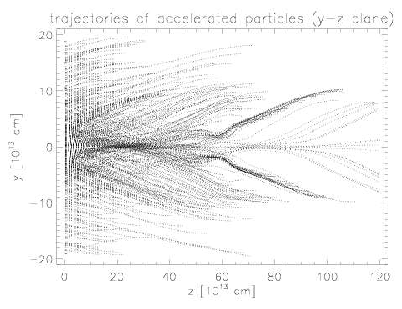,width=0.8\textwidth}
\caption[]{}
\end{figure}
\begin{figure}
    \psfig{file=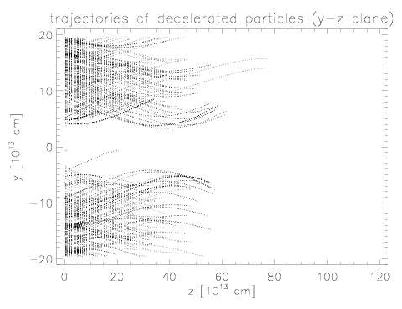,width=0.8\textwidth}
\caption[]{}
\end{figure}

\begin{figure}
    \psfig{file=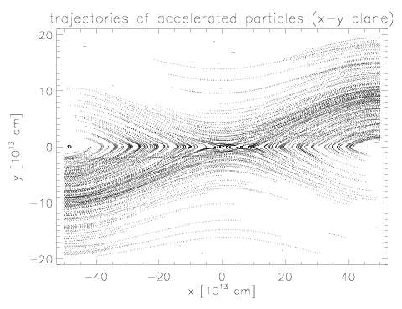,width=0.8\textwidth}
\caption[]{}
\end{figure}
\begin{figure}
    \psfig{file=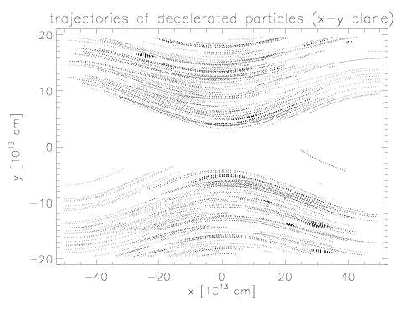,width=0.8\textwidth}
\caption[]{}
\end{figure}

\begin{figure}
    \psfig{file=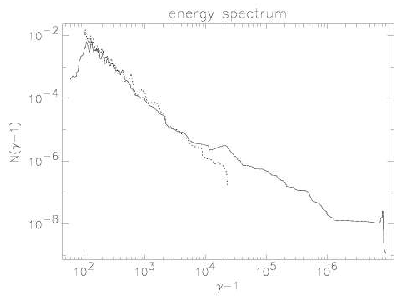,width=0.8\textwidth}
\caption[]{}
\end{figure}

\begin{figure}
    \psfig{file=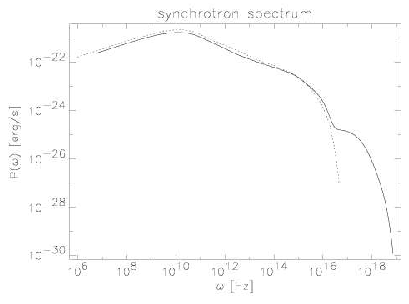,width=0.8\textwidth}
\caption[]{}
\end{figure}
\clearpage

\begin{figure}
    \psfig{file=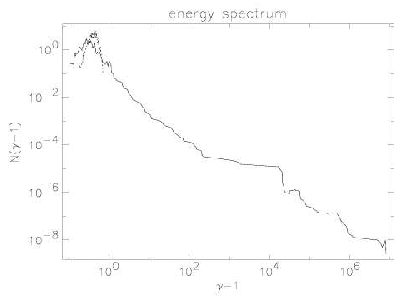,width=0.8\textwidth}
\caption[]{}
\end{figure}

\begin{figure}
    \psfig{file=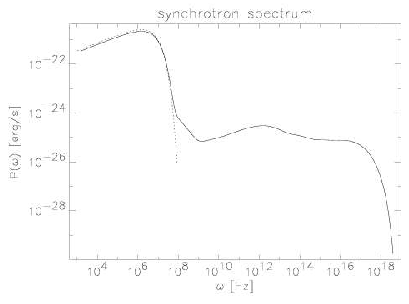,width=0.8\textwidth}
\caption[]{}
\end{figure}
\clearpage

\end{document}